\documentclass[pdftex,aps,prl,amsmath,amsfonts,amssymb,superscriptaddress,twocolumn,showpacs]{revtex4}
\renewcommand{\vec}[1]{\mathbf{#1}}
\usepackage[pdftex]{graphicx}
\usepackage{wasysym}
\usepackage{setspace}
\usepackage{color}
\usepackage{nicefrac}
\usepackage{times}

\renewcommand{\Re}{\operatorname{Re}}
\renewcommand{\Im}{\operatorname{Im}}

\renewcommand{\d}{\mathrm{d}}

\newcommand{\figref}[1]{Fig.~\ref{fig:#1}}

\renewcommand{\eqref}[1]{Eq.~(\ref{eq:#1})}

\newcommand{\citeasnoun}[1]{Ref.~\onlinecite{#1}}

\def\a{s}
\def\b{s}
\newcommand{\add}[1]{\if\a\b{{\color{red} #1}}\else{#1}\fi}
\newcommand{\comm}[1]{\if\a\b{{\color{blue}\{\small \sc #1\}}}\else{}\fi}
\newcommand{\del}[1]{{\if\a\b{{\color{magenta}[[#1]]}}\else{}\fi}}

\begin{document}

\title{Anomalous near-field heat transfer between a cylinder and a perforated surface}

\author{Alejandro W. Rodriguez}
\affiliation{School of Engineering and Applied Sciences, Harvard University, Cambridge, MA 02138}
\affiliation{Department of Mathematics, Massachusetts Institute of Technology, Cambridge, MA 02139}
\author{M. T. Homer Reid}
\affiliation{Department of Mathematics, Massachusetts Institute of Technology, Cambridge, MA 02139}
\author{Jaime Varela}
\affiliation{Department of Physics, University of California Berkeley, Berkeley, CA 94704}
\author{John~D.~Joannopoulos}
\affiliation{Department of Physics, Massachusetts Institute of Technology, Cambridge, MA 02139}
\author{Federico Capasso}
\affiliation{School of Engineering and Applied Sciences, Harvard University, Cambridge, MA 02138}
\author{Steven G. Johnson}
\affiliation{Department of Mathematics, Massachusetts Institute of Technology, Cambridge, MA 02139}

\begin{abstract}
  We predict that the radiative heat-transfer rate between a cylinder
  and a perforated surface depends \emph{non}monotonically on their
  separation.  This anomalous behavior, which arises due to near-field
  effects, is explained using a heuristic model based on the
  interaction of a dipole with a plate. We show that nonmonotonicity
  depends not only on geometry and temperature but also on material
  dispersion---for micron and submicron objects, nonmonotonicity is
  present in polar dielectrics but absent in metals with small skin
  depths.
\end{abstract}

\maketitle



Disconnected bodies of different temperatures can exchange energy
through stochastic electromagnetic
waves~\cite{Rytov89,Mulet02,Joulain05,Carey06,Volokitin07,Zhang07,BasuZhang09},
a phenomenon known as radiative heat transfer that underlies many
naturally occurring and technologically relevant
processes~\cite{BasuZhang09}. Recent advances in microfabrication and
metrology have enabled experiments that can now regularly probe this
phenomenon at micron and submicron
scales~\cite{Rousseau09,Sheng09}. At such small scales, unusual
near-field interactions arise~\cite{BasuZhang09}, but non-planar
geometries in this context are only just beginning to be
explored~\cite{Narayanaswamy08:spheres,bimonte09,Messina11,RodriguezIl11,
  Kruger11,OteyFan11,McCauleyReid12,Lussange12,Guerout12,RodriguezReid12}. In
this Letter, inspired by our previous work on Casimir
repulsion~\cite{LevinMc10}, we demonstrate that the heat transfer
between a cylinder (or elongated object) and a perforated surface
(e.g. a ring) can vary \emph{nonmonotonically} with respect to their
mutual separation, in contrast to what has been observed in all
previous geometries~\cite{BasuZhang09}. This anomalous effect stems
primarily from the contribution of dipolar (near) fields, as we
illustrate by a heuristic model in which the cylinder is modeled as a
quasi-static dipole and the ring as an infinitesimally thin plate with
a hole. We find that nonmonotonicity weakens (and eventually
disappears) whenever the geometrical and material parameters of the
objects deviate significantly from the dipolar regime: for cylinders
of equal or nearly equal aspect ratio, large ring thicknesses, large
temperatures, or metals with small skin depths (such as gold), the
usual monotonic dependence is observed. We note that, in contrast to
conventional geometries, this effect cannot be predicted even
qualitatively by ``additive'' approximations such as the well-known
proximity approximation.

In the far field (object separations $d$ much greater than the thermal
wavelength $\lambda_T = \hbar c / k_B T$), radiative heat transfer is
dominated by the exchange of propagating waves and is thus nearly
insensitive to changes in separations.  In the (less studied) near
field (object separations $d\lesssim \lambda_T$), not only are
interference effects important, but otherwise-negligible evanescent
waves also contribute flux~\cite{Zhang07,BasuZhang09}. Such near-field
effects have been most commonly studied in planar geometries, where
they are known to lead to monotonically increasing heat transfer rates
with decreasing $d$, resulting in orders-of-magnitude enhancements of
the total heat transfer (which can even exceed the far-field
black-body limit at sub-micron separations~\cite{BasuZhang09}). Thus
far, little is known about the heat transfer characteristics of bodies
whose shapes differ significantly from the planar, unpatterned
structures of the past. Recent theoretical progress include
predictions for a handful of new geometries, including
spheres~\cite{Narayanaswamy08:spheres,OteyFan11} and
cones~\cite{McCauleyReid12} suspended above slabs, as well as
patterned
surfaces~\cite{bimonte09,Messina11,Kruger11,RodriguezIl11,Lussange12}. Our
work extends these studies to yet another class of possible
geometries: interleaved bodies whose near-field interactions and
shapes lead to anomalous heat-transport phenomena.

The heat transfer rate $H$ between two objects held at temperatures
$T_1$ and $T_2$ can be expressed in the
form~\cite{Zhang07,BasuZhang09}:
\begin{equation}
  H = \int_{0}^\infty d\omega \left[\Theta(\omega,T_1) -
    \Theta(\omega,T_2)\right] \Phi(\omega),
\end{equation}
where $\Phi$ is the \emph{flux spectrum} (the time-averaged flux into
object 2 due to current sources in object 1), and $\Theta(\omega,T) =
\hbar \omega / [\exp(\hbar \omega / k_B T) - 1]$ is the mean Planck
energy per oscillator at frequency $\omega$ and temperature $T$. (Note
that $\Phi=1$ for black bodies that capture all of one another's
radiation.) We compute $\Phi$ by exploiting two recent computational
methods: a fluctuating surface-current (FSC) formulation involving the
solution of an integral equation at each
frequency~\cite{RodriguezReid12}, and a Langevin finite-difference
time-domain (FDTD) formulation in which one explicitly time-evolves
Maxwell's equations in response to (broad-bandwidth) stochastic
sources inside the bodies~\cite{RodriguezIl11}. In order to
distinguish the effects of geometry from those of material dispersion,
we begin by considering a simple model material: a lossy dielectric
with a broad (low-dispersion) absorption peak, given by
$\varepsilon(\omega) = \varepsilon_\infty - \sigma / (\omega_0^2 -
\omega^2 - i\gamma \omega)$, with $\varepsilon_\infty = 12.5$, $\sigma
= 4\times 10^2~(c/\mu\mathrm{m})^2$, $\omega_0 = 0$, and $\gamma =
60~(c/\mu\mathrm{m})$, corresponding to roughly-constant $\Re
\varepsilon \approx 12$ and large $\Im \varepsilon \gtrsim 1$ over
relevant frequencies. Later, we consider realistic materials and show
that material dispersion also plays a crucial role.

\begin{figure}[t]
\centering
\includegraphics[width=1.0\columnwidth]{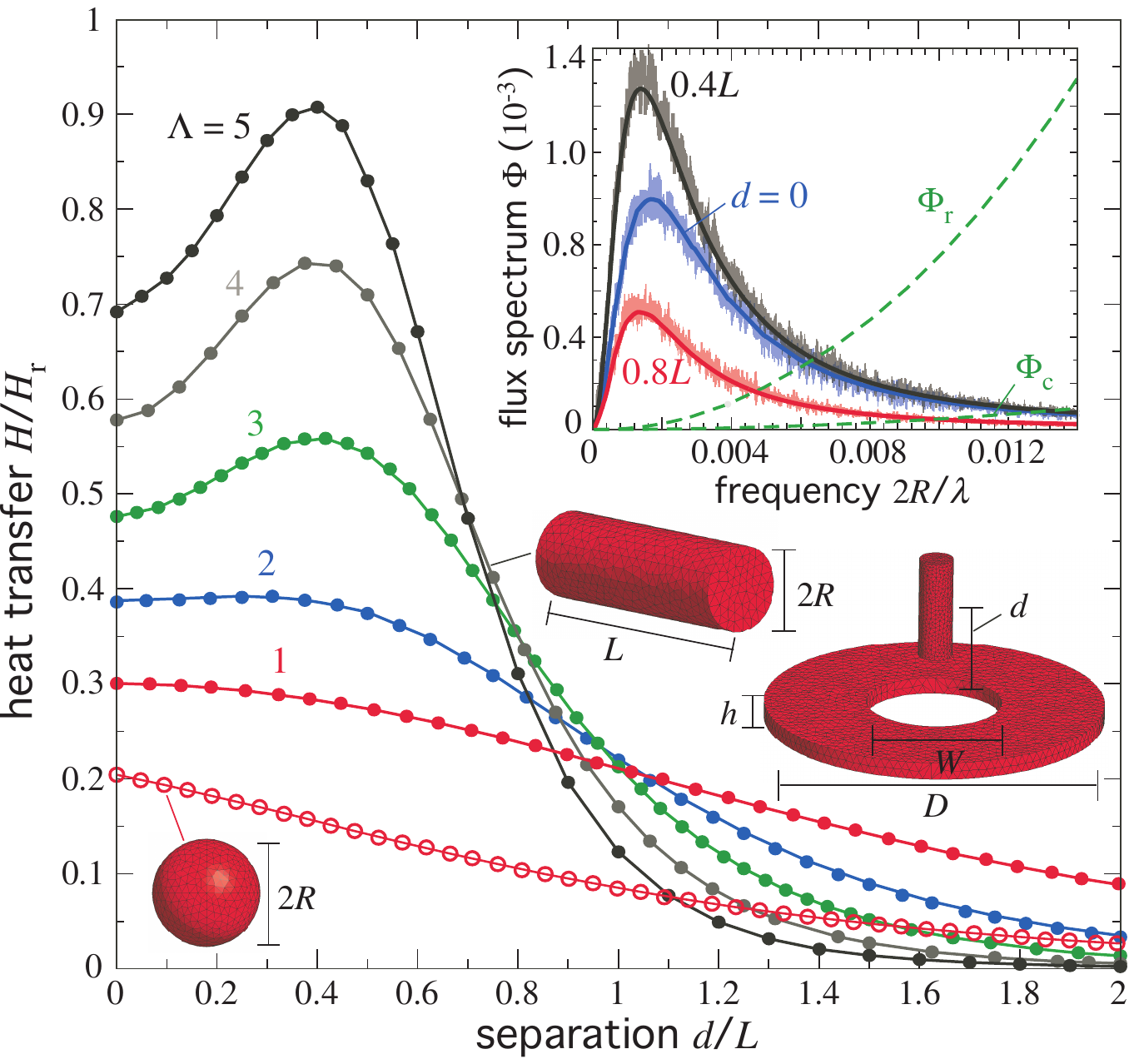}
\caption{Heat-transfer rate $H$ from a room temperature ring to a
  cylinder (or sphere) of fixed radius $R=0.1\mu$m and aspect ratio
  $\Lambda=L/2R$ held at $T=0$, as a function of their center--center
  separation $d$. $H$ is normalized by the heat radiation of the
  isolated ring $H_\mathrm{r}$. Both objects are lossy dielectrics
  with $\Re \varepsilon \approx 12$ (see text).  Inset shows the flux
  spectrum $\Phi(\omega)$ of the $\Lambda=5$ configuration at three
  separations, and also of the isolated cylinder and ring (dashed
  lines).}
\label{fig:fig1}
\end{figure}

We previously studied the radiation of \emph{isolated} cylinders and
rings~\cite{RodriguezReid12}. In this work, we consider the new
phenomena that arise when these two objects are brought into close
proximity so that near-field effects are present.  The dashed green
lines in the inset of \figref{fig1} show the flux spectrum of an
isolated cylinder ($\Phi_\mathrm{c}$) of radius $R=0.1\mu$m and aspect
ratio $\Lambda = L/2R = 5$, and of an isolated ring
($\Phi_\mathrm{r}$) of outer (inner) diameter $D=2\mu$m ($W =
0.8\mu$m) and thickness $h=0.05D$, as a function of frequency (units
of $2R/\lambda$). In this long-wavelength limit ($\lambda \gg R$),
large $\Im \varepsilon$ and the absence of geometric and material
resonances means that both objects emit significantly less than an
ideal black body, so that $\Phi_c \ll \Phi_r \ll 1$.

The inset in \figref{fig1} also shows the flux spectrum $\Phi$ when
the two objects are brought into close proximity (solid lines), at
three different center--center separations $d$, showing dramatic
changes from the isolated case. $\Phi$ is computed via both FDTD
(noisy curves) and FSC (smooth curves) methods, showing excellent
agreement; the remaining calculations use FSC only. Compared to
isolated objects, the increasing contribution of evanescent waves in
both objects leads to an overall increase in the flux at low
frequencies. [Note that the peak in $\Phi$ at $\lambda \approx 10^{-3}
  R$ is a consequence of material dispersion: the loss tangent of the
  material $\sim \Im \varepsilon / \Re \varepsilon \to \infty$ as
  $\lambda \to \infty$, leading to zero radiation; the peak in the
  spectrum occurs at the cross-over wavelength for which $\Im
  \varepsilon \sim \Re \varepsilon$. Low-frequency cut-offs in the
  near-field enhancement occur in highly conductive materials, such as
  gold (below).] Most interestingly, however, the enhancement in
$\Phi$ here does not increase monotonically with decreasing $d$:
$\Phi$ \emph{increases} from $d=\infty$ to $d\approx 0.4L$, but
\emph{decreases} from $d\approx 0.4L$ to $d=0$.

\begin{figure}[t]
\centering
\includegraphics[width=1.0\columnwidth]{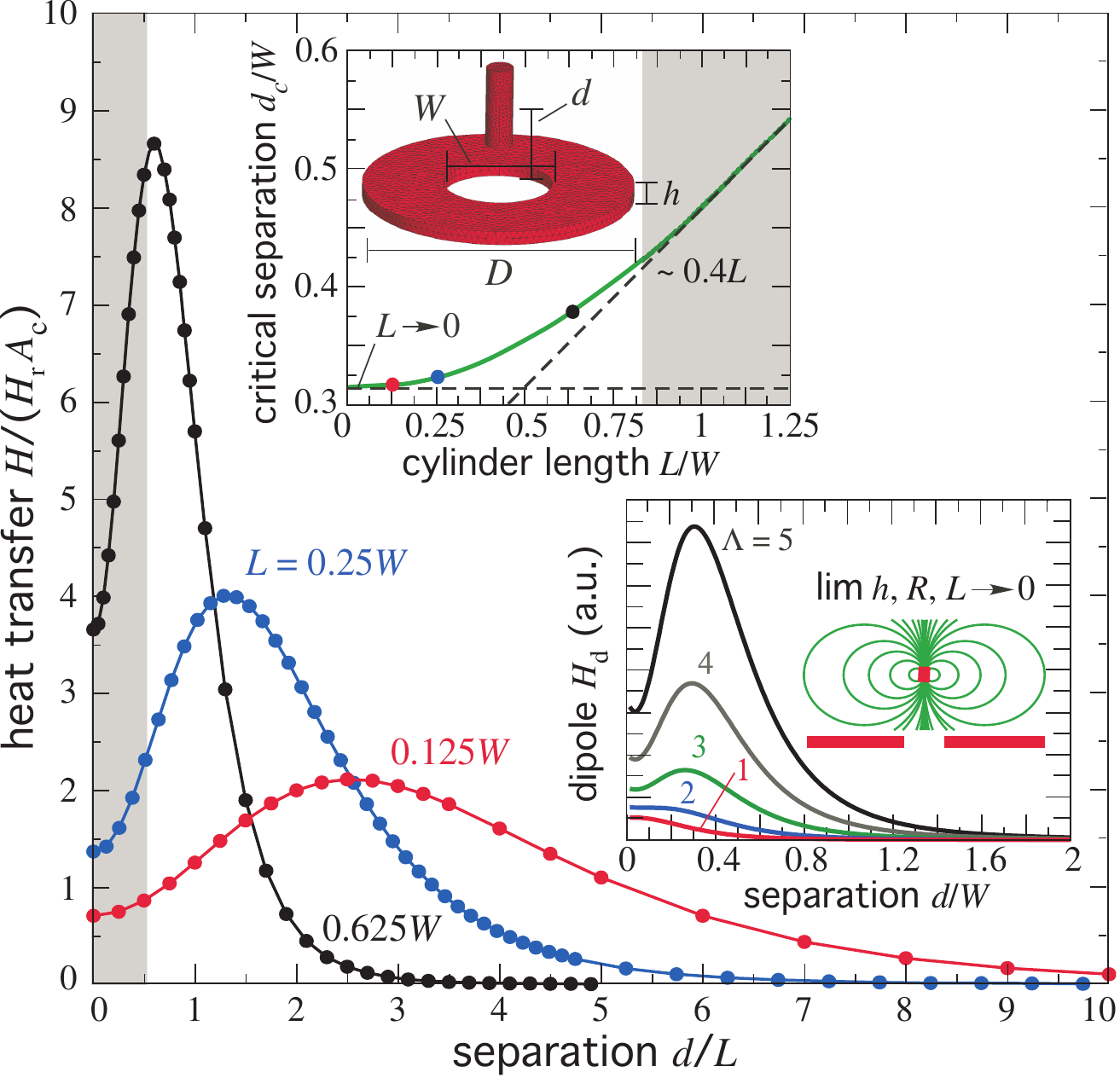}
\caption{Heat-transfer rate $H$ between the ring and cylinder of
  \figref{fig1}, for fixed cylinder aspect ratio $\Lambda=5$ and thin
  ring thickness $h=5\times 10^{-3}D$, as a function of $d$ and for
  multiple cylinder lengths $L$. $H$ is normalized by the radiation of
  the isolated ring $H_\mathrm{r}$ multiplied by the cylinder area
  $A_\mathrm{c}$. The shaded region denotes separations over which the
  two objects are interleaved. Top inset shows the critical separation
  $d_c$ of largest $H$ as a function of $L$, with shaded areas again
  denoting interleaved configurations. Bottom inset shows $H$ in the
  limit $h,R,L\to 0$, as computed by a heuristic model (see text).}
\label{fig:fig2}
\end{figure}

To explore the geometry dependence of this near-field behavior, we now
examine the overall heat-transfer rate $H$ as a function of $d$
instead of the spectrum. In particular, \figref{fig1} shows $H$ from a
room-temperature ring to a cylinder at $T=0$, for multiple aspect
ratios $\Lambda$. For comparison, $H$ is normalized to the radiation
rate of the isolated ring $H_\mathrm{r}$. For large anisotropy
$\Lambda=5$, $H$ first \emph{increases} as the two objects approach
each other due to the usual near-field enhancement, and then decreases
as $d \to 0$, peaking at a critical separation $d_c \approx 0.4L$.
Unlike previously studied structures involving non-interleaved
objects, the heat transfer in this geometry does not diverge as $d \to
0$: although the two objects approach each other in this limit, they
never touch. Also, $H \to 0$ as $d \to \infty$ due to the finite size
of the two objects. As $\Lambda$ decreases (keeping $R$ fixed),
corresponding to increasingly isotropic cylinders, the nonmonotonicity
becomes less pronounced, and is completely absent in both the
$\Lambda=1$ configuration (small anisotropy) and for a sphere (open
red circles). As expected, there is an overall decrease in $H$ with
decreasing $L$ due to the decreasing volume of the
cylinder. Nonmonotonicity also slowly disappears as the ring thickness
$h$ increases to $h \approx 0.5L$, leaving a relatively wide range of
thicknesses over which the effect can be observed. Furthermore, as
expected, the strength of the nonmonotonicity $H(d_c) / H(0)$ grows
larger as the cylinder surface approaches the rim of the ring
(corresponding to larger $R$ or smaller $W$) due to near-field
effects, and also for larger $D$ due to the larger surface area. More
interesting however, is the fact that nonmonotonicity persists even
when the cylinders are shifted laterally (shifts $\lesssim 0.5W$), an
asymmetric configuration that is likely to occur in experiments.
Finally, we find that $H(d_c) / H(0)$ increases as $h, R, L \to 0$
(for fixed $\Lambda$), as seen below (\figref{fig2}).

Cylindrical symmetry allows us to decompose $\Phi$ into azimuthal
angular components $m$ (fields $\sim e^{im\theta}$), implemented in
FDTD with cylindrical coordinates. Our calculations reveal (not shown)
that most (though not all) of the nonmonotonic dependence comes from
the contribution of dipolar ($m=0$) fields, which dominate the heat
transfer at these long thermal wavelengths $\lambda_T \gg R,L$. At
these wavelengths, a cylinder with large $\Lambda$ will act like a
fluctuating dipole oriented mainly along the symmetry axis. Since the
fields generated by a fluctuating dipole are polarized mostly along
the dipole axis and since current fluctuations in the thin ring are
polarized mostly along the plane of the ring, it follows that the
fields induced by a long cylinder will do less work on currents in the
ring whenever the objects are nearly co-planar ($d \to 0$).

We quantify this argument by focusing on a simple (albeit heuristic)
model in which the cylinder is modeled as a dipole of electric
polarizability $\boldsymbol{\alpha}^E(\omega)$ and the ring as an
infinitesimally thin plate with a hole. For convenience, we only
consider the non-retarded (quasi-static) limit of small dipole
separations $d \ll \lambda_T$, in which case the heat transfer rate
between the dipole and plate can be expressed as $H_{\d} = \int
d\omega \left[\Theta(\omega,T_1) - \Theta(\omega,T_2)\right]
\Phi_{\d}(\omega)$, where the flux spectrum is given
by~\cite{Volokitin07, Chapuis08prb, Huth10}:
\begin{equation}
  \Phi_{\d}(\omega) = \frac{2}{\pi} \sum_i \omega^2
  \Im[\alpha^E_{i}(\omega)] \, \Im \left[ G_{ii}(\vec{r}_\d,\vec{r}_\d)
    \right],
\label{eq:phid}
\end{equation}
Here, $G_{ij}(\vec{r},\vec{r}')$ is the electric Green's function of
the plate---the electric field in the $i$th direction at $\vec{r}$ due
to a dipole source in the $j$th direction at $\vec{r}'$---and
$\vec{r}_\d$ is the location of the dipole. The calculation of
$G_{ij}$ for a perfectly conducting, infinitely thin plate was carried
out in~\citeasnoun{Eberlein11} for the purpose of computing the
Casimir-Polder force in that idealized system. However, because perect
conductors do not radiate ($\Im G_{ij} = 0$), we merely exploit that
expression as the starting point of a quasi-static perturbative
calculation in which the plate is assumed to have a \emph{small amount
  of absorption}. In particular, we are interested in computing the
dissipated power or Ohmic losses on a plate with small but finite
conductivity $\sigma$, given the quasi-static fields at the surface of
the perfectly conducting plate. Following~\citeasnoun{Jackson98}, the
resistive losses on the plate are $\sim \Im G_{jj}(\vec{r}_\d,
\vec{r}_\d) \sim \int d^2\vec{r} \, \sigma(\vec{r})
|\vec{G}^0_{j}(\vec{r},\vec{r}_\d)|^2$, where $G^0_{ij}$ is the
electric Dyadic Green's function of the unperturbed (perfectly
conducting) plate, and the integral is performed over the plate
surface.  It turns out that $G^0$ exhibits non-integrable
singularities at the rim of the hole, a well-known artifact of the
idealized nature of corners and wedges in
electromagnetism~\cite{Jackson98}. While the electromagnetic energy
corresponding to these fields is finite when integrated over all
space, the singularity in the fields is problematic for the
perturbation theory since the form of the perturbation considered here
requires that $G^0$ be integrated only over the plane of the
plate. Essentially, this model does not account for the finite
thickness of the plate, and consequently fails to capture effects
associated with the finite penetration or skin-depth $\delta =
c/(\omega \Im \sqrt{\varepsilon})$ of fields. Therefore, we use
$\delta$ as a cutoff lengthscale to regularize the integral near the
rim of the hole. Surprisingly, and despite its many shortcomings, this
heuristic model captures most of the features of interest.

The bottom inset in \figref{fig2} shows this heuristic model's heat
transfer rate $H_\d$ (in arbitrary units) from a room-temperature
plate of conductivity $\sigma=\omega \Im \varepsilon$, infinitesimally
small thickness $h\to 0$, and hole diameter $W$, to a small cylinder
of radius $R \ll W$ and electric polarizability
$\boldsymbol{\alpha}^E$ held at $T=0$, as a function of their
separation $d$. $H_\d$ is computed by \eqref{phid} using the dipole
model above, with the polarizability of the dipole taken to be that of
a uniform prolate spheroid~\cite{Huth10} of aspect ratio $\Lambda =
L/2R$ and permittivity $\varepsilon$ (same as above). For $\Lambda =
5$, corresponding to a highly anisotropic object, one observes the
expected nonmonotonic behavior, with the critical separation $d_c
\approx 0.3W$ now determined by $W$. As before, nonmonotonicity
decreases with decreasing $\Lambda$, disappearing completely in the
limit $\Lambda \to 1$ of an isotropic (spherical) object with
polarizability $\alpha^E = 4/3\pi R^3 (\varepsilon - 1) / (\varepsilon
+ 2)$. In comparison with \figref{fig1}, we observe that the simple
model quantitatively captures the onset of nonmonotonicity at $\Lambda
\gtrsim 2$. Since the model represents a point-dipole limit, we also
compare to exact calculations for fixed $\Lambda=5$ by letting $L\to
0$ and $h=5\times 10^{-3}D$, as shown in \figref{fig2}. While
nonmonotonicity is present for all $L$, the scaling of the critical
separation $d_c$ with $L$ changes qualitatively as $L \to
0$. Specifically, as shown by the top inset of \figref{fig2}, $d_c
\approx 0.3W$ for $L \ll W$, in quantitative agreement with the dipole
model, while $d_c \approx 0.4L$ for $L \gg W$, with the cross-over
regime occurring for $L \approx 0.5W$. We find that for $L \lesssim
0.8W$, the onset of nonmonotonicity ($d_c$) occurs before the two
objects are interleaved, i.e. when there is a \emph{separating plane}
between the objects.

\begin{figure}[t]
\includegraphics[width=1.0\columnwidth]{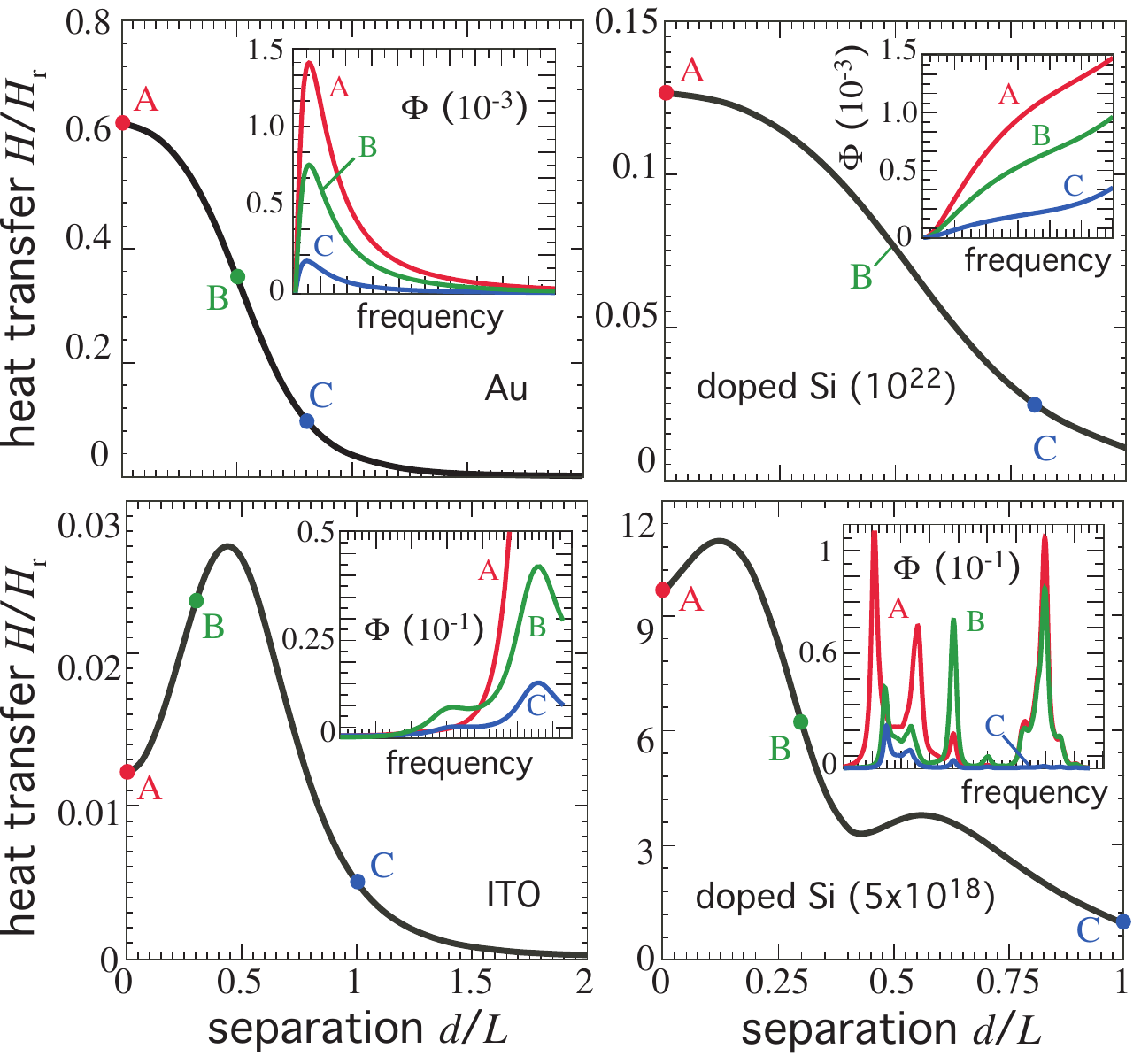}
\caption{Heat transfer rate $H$ between the ring and cylinder of
  \figref{fig1}, for fixed cylinder aspect ratio $\Lambda=5$, as a
  function of separation $d$. $H$ is normalized to the isolated
  cylinder radiation rate $H_\mathrm{r}$, and plotted for multiple
  material configurations. Insets show the corresponding flux spectra
  $\Phi(\omega)$ at three different (labeled) separations.}
\label{fig:fig3}
\end{figure}

Aside from geometry, changes coming from either temperature or
material dispersion leading to deviations from the ideal dipole regime
can also weaken nonmonotonic behavior. Thus far, we have restricted
ourselves to studying micro-scale bodies near room temperature (which
emit preferentially at infrared frequencies), corresponding to large
thermal wavelengths $\lambda_T \gg R$. At larger temperatures
($\lambda_T \lesssim R$) however, such bodies can no longer be well
described as dipole emitters, and thus $\Phi$ no longer exhibits
nonmonotonic behavior. We find that nonmonotonicity persists at
temperatures well beyond the mere $T\approx 300$~K considered here, so
long as the feature sizes of the objects involved remain at or below
the micron scale. Realistic materials often exhibit substantial
material dispersion at or near infrared wavelengths, and this can also
significantly alter the dipole picture above. This situation is
depicted in \figref{fig3}, which shows $H$ for various material
configurations, including metals (gold and indium tin oxide) and polar
dielectrics (doped silicon). (The Au and ITO dispersions are
determined by Drude models with plasma frequencies $\omega_p = 1.367
\times 10^{16}$~rad/s and $\omega_p = 1.4739\times 10^{15}$~rad/s, and
relaxation rates $\gamma = 5.317\times 10^{13}$~rad/s and $\gamma =
1.5347\times 10^{14}$~rad/s, respectively, whereas the doped silicon
dispersion is given by the model of~\citeasnoun{Duraffourg06}.)
Interestingly, we find that nonmonotonicity is completely absent for
the Au and highly-doped silicon ($10^{22}$cm$^{-3}$) configurations,
and present in the ITO and lesser-doped silicon ($5\times
10^{18}$cm$^{-3}$) configurations. The reason for the discrepancy
comes from the fact that highly conductive metals and polar
dielectrics respond very differently to incident light. In particular,
for metals, the electric dipole approximation (above) breaks down for
skin depths $\delta \ll R,L$: in that limit, eddy currents induced on
the surface of the metallic objects also lead to large \emph{magnetic}
dipole moments~\cite{Chapuis08prb, Huth10}, with magnetic
polarizabilities $\boldsymbol{\alpha}^H \gg
\boldsymbol{\alpha}^E$. Unfortunately, the interaction between a
magnetic dipole moment and a plate does not exhibit the desired
nonmonotonic effect (at least in this geometry), which explains the
results in the case of Au and highly-doped silicon cylinders, whose
skin depths $\delta \approx 10^{-2}\mu\mathrm{m} \ll R$ at infrared
wavelengths. If we scale the entire structure down to much smaller
scales $R,L \ll \delta$ (not shown), we find that nonmonotonicity is
restored.  It follows that for highly conductive materials, one
obtains the desired nonmonotonic effect only for $h,R,L \ll \delta \ll
\lambda_T$.

The insets in \figref{fig3} show the flux spectra of the various
material configurations at three separations (labeled $A$, $B$, and
$C$). The flux spectrum for Au and highly-doped silicon are monotonic
with $d$ at all frequencies. On the other hand, ITO and lesser-doped
silicon are nonmonotonic with $d$ only at certain frequencies. For any
material, $\Phi$ is monotonic at high frequencies ($\lambda \ll R,L$),
where the cylinder no longer acts as a dipole, but this transition
occurs even more rapidly for ITO (which is only nonmonotonic for low
frequencies) because of the aforementioned skin-depth effect. For
silicon, the situation is greatly complicated by the presence of
multiple \emph{geometric} resonances ($\varepsilon$ has a single
absorption peak), arising from the ability of the fields to probe the
interior as well as the surface of the dielectric. It turns out that
only two of the resonant peaks in this case exhibit nonmonotonicity
with $d$, corresponding to resonances with the necessary dipole-like
polarizations.

Similar and even more pronounced nonmonotonic behaviors should arise
in other geometries, so long as the suspended objects (regardless of
shape) are sufficiently anisotropic (behave dipole-like) and radiate
primarily in the direction orthogonal to the patterned surface.  An
interesting structure to explore in the future is a nanowire array
suspended above a periodically patterned thin film.

This work was supported by DARPA Contract No. N66001-09-1-2070-DOD and
by the AFOSR Multidisciplinary Research Program of the University
Research Initiative (MURI) for Complex and Robust On-chip
Nanophotonics, Grant No. FA9550-09-1-0704.


\begin{thebibliography}{27}
\expandafter\ifx\csname natexlab\endcsname\relax\def\natexlab#1{#1}\fi
\expandafter\ifx\csname bibnamefont\endcsname\relax
  \def\bibnamefont#1{#1}\fi
\expandafter\ifx\csname bibfnamefont\endcsname\relax
  \def\bibfnamefont#1{#1}\fi
\expandafter\ifx\csname citenamefont\endcsname\relax
  \def\citenamefont#1{#1}\fi
\expandafter\ifx\csname url\endcsname\relax
  \def\url#1{\texttt{#1}}\fi
\expandafter\ifx\csname urlprefix\endcsname\relax\def\urlprefix{URL }\fi
\providecommand{\bibinfo}[2]{#2}
\providecommand{\eprint}[2][]{\url{#2}}

\bibitem[{\citenamefont{Rytov et~al.}(1989)\citenamefont{Rytov, Tatarskii, and
  Kravtsov}}]{Rytov89}
\bibinfo{author}{\bibfnamefont{S.~M.} \bibnamefont{Rytov}},
  \bibinfo{author}{\bibfnamefont{V.~I.} \bibnamefont{Tatarskii}},
  \bibnamefont{and} \bibinfo{author}{\bibfnamefont{Y.~A.}
  \bibnamefont{Kravtsov}}, \emph{\bibinfo{title}{Principles of Statistical
  Radiophsics II: Correlation Theory of Random Processes}}
  (\bibinfo{publisher}{Springer-Verlag}, \bibinfo{year}{1989}).

\bibitem[{\citenamefont{Mulet et~al.}(2002)\citenamefont{Mulet, Joulain,
 Carminati, and Greffet}}]{Mulet02}
\bibinfo{author}{\bibfnamefont{J.-P.} \bibnamefont{Mulet}},
  \bibinfo{author}{\bibfnamefont{K.}~\bibnamefont{Joulain}},
  \bibinfo{author}{\bibfnamefont{R.}~\bibnamefont{Carminati}},
  \bibnamefont{and} \bibinfo{author}{\bibfnamefont{J.-J.}
  \bibnamefont{Greffet}}, \bibinfo{journal}{Micro. Thermophys. Eng.}
  \textbf{\bibinfo{volume}{6}}, \bibinfo{pages}{209} (\bibinfo{year}{2002}).

\bibitem[{\citenamefont{Joulain et~al.}(2005)\citenamefont{Joulain, Mulet,
  Marquier, Carminati, and Greffet}}]{Joulain05}
\bibinfo{author}{\bibfnamefont{K.}~\bibnamefont{Joulain}},
  \bibinfo{author}{\bibfnamefont{J.-P.} \bibnamefont{Mulet}},
  \bibinfo{author}{\bibfnamefont{F.}~\bibnamefont{Marquier}},
  \bibinfo{author}{\bibfnamefont{R.}~\bibnamefont{Carminati}},
  \bibnamefont{and} \bibinfo{author}{\bibfnamefont{J.-J.}
  \bibnamefont{Greffet}}, \bibinfo{journal}{Surf. Sci. Rep.}
  \textbf{\bibinfo{volume}{57}}, \bibinfo{pages}{59} (\bibinfo{year}{2005}).

\bibitem[{\citenamefont{Carey et~al.}(2006)\citenamefont{Carey, Cheng,
  Grigoropoulos, Kaviany, and Majumdar}}]{Carey06}
\bibinfo{author}{\bibfnamefont{V.~P.} \bibnamefont{Carey}},
  \bibinfo{author}{\bibfnamefont{G.}~\bibnamefont{Cheng}},
  \bibinfo{author}{\bibfnamefont{C.}~\bibnamefont{Grigoropoulos}},
  \bibinfo{author}{\bibfnamefont{M.}~\bibnamefont{Kaviany}}, \bibnamefont{and}
  \bibinfo{author}{\bibfnamefont{A.}~\bibnamefont{Majumdar}},
  \bibinfo{journal}{Nanoscale Micro. Thermophys. Eng.}
  \textbf{\bibinfo{volume}{12}}, \bibinfo{pages}{1} (\bibinfo{year}{2006}).

\bibitem[{\citenamefont{Volokitin and Persson}(2007)}]{Volokitin07}
\bibinfo{author}{\bibfnamefont{A.~I.} \bibnamefont{Volokitin}}
  \bibnamefont{and} \bibinfo{author}{\bibfnamefont{B.~N.~J.}
  \bibnamefont{Persson}}, \bibinfo{journal}{Rev. Mod. Phys.}
  \textbf{\bibinfo{volume}{79}}, \bibinfo{pages}{1291} (\bibinfo{year}{2007}).

\bibitem[{\citenamefont{Zhang}(2007)}]{Zhang07}
\bibinfo{author}{\bibfnamefont{Z.~M.} \bibnamefont{Zhang}},
  \emph{\bibinfo{title}{Nano/Microscale Heat Transfer}}
  (\bibinfo{publisher}{McGraw-Hill}, \bibinfo{address}{New York},
  \bibinfo{year}{2007}).

\bibitem[{\citenamefont{Basu et~al.}(2009)\citenamefont{Basu, Zhang, and
  Fu}}]{BasuZhang09}
\bibinfo{author}{\bibfnamefont{S.}~\bibnamefont{Basu}},
  \bibinfo{author}{\bibfnamefont{Z.~M.} \bibnamefont{Zhang}}, \bibnamefont{and}
  \bibinfo{author}{\bibfnamefont{C.~J.} \bibnamefont{Fu}},
  \bibinfo{journal}{Int. J. Energy Res.} \textbf{\bibinfo{volume}{33}},
  \bibinfo{pages}{1203} (\bibinfo{year}{2009}).

\bibitem[{\citenamefont{Rousseau et~al.}(2009)\citenamefont{Rousseau, Siria,
  Guillaume, Volz, Comin, Chevrier, and Greffet}}]{Rousseau09}
\bibinfo{author}{\bibfnamefont{E.}~\bibnamefont{Rousseau}},
  \bibinfo{author}{\bibfnamefont{A.}~\bibnamefont{Siria}},
  \bibinfo{author}{\bibfnamefont{J.}~\bibnamefont{Guillaume}},
  \bibinfo{author}{\bibfnamefont{S.}~\bibnamefont{Volz}},
  \bibinfo{author}{\bibfnamefont{F.}~\bibnamefont{Comin}},
  \bibinfo{author}{\bibfnamefont{J.}~\bibnamefont{Chevrier}}, \bibnamefont{and}
  \bibinfo{author}{\bibfnamefont{J.-J.} \bibnamefont{Greffet}},
  \bibinfo{journal}{Nat. Phot.} \textbf{\bibinfo{volume}{3}},
  \bibinfo{pages}{514} (\bibinfo{year}{2009}).

\bibitem[{\citenamefont{Shen et~al.}(2009)\citenamefont{Shen, Narayanaswamy,
  and Chen}}]{Sheng09}
\bibinfo{author}{\bibfnamefont{S.}~\bibnamefont{Shen}},
  \bibinfo{author}{\bibfnamefont{A.}~\bibnamefont{Narayanaswamy}},
  \bibnamefont{and} \bibinfo{author}{\bibfnamefont{G.}~\bibnamefont{Chen}},
  \bibinfo{journal}{Nano Letters} \textbf{\bibinfo{volume}{9}},
  \bibinfo{pages}{2909} (\bibinfo{year}{2009}).

\bibitem[{\citenamefont{Narayanaswamy and
  Chen}(2008)}]{Narayanaswamy08:spheres}
\bibinfo{author}{\bibfnamefont{A.}~\bibnamefont{Narayanaswamy}}
  \bibnamefont{and} \bibinfo{author}{\bibfnamefont{G.}~\bibnamefont{Chen}},
  \bibinfo{journal}{Phys. Rev. B} \textbf{\bibinfo{volume}{77}},
  \bibinfo{pages}{075125} (\bibinfo{year}{2008}).

\bibitem[{\citenamefont{Messina and Antezza}(2011)}]{Messina11}
\bibinfo{author}{\bibfnamefont{R.}~\bibnamefont{Messina}} \bibnamefont{and}
  \bibinfo{author}{\bibfnamefont{M.}~\bibnamefont{Antezza}},
  \bibinfo{journal}{Phys. Rev.~A} \textbf{\bibinfo{volume}{84}},
  \bibinfo{pages}{042102} (\bibinfo{year}{2011}).

\bibitem[{\citenamefont{Rodriguez et~al.}(2011)\citenamefont{Rodriguez, Ilic,
  Bermel, Celanovic, Joannopoulos, Soljacic, and Johnson}}]{RodriguezIl11}
\bibinfo{author}{\bibfnamefont{A.~W.} \bibnamefont{Rodriguez}},
  \bibinfo{author}{\bibfnamefont{O.}~\bibnamefont{Ilic}},
  \bibinfo{author}{\bibfnamefont{P.}~\bibnamefont{Bermel}},
  \bibinfo{author}{\bibfnamefont{I.}~\bibnamefont{Celanovic}},
  \bibinfo{author}{\bibfnamefont{J.~D.} \bibnamefont{Joannopoulos}},
  \bibinfo{author}{\bibfnamefont{M.}~\bibnamefont{Soljacic}}, \bibnamefont{and}
  \bibinfo{author}{\bibfnamefont{S.~G.} \bibnamefont{Johnson}},
  \bibinfo{journal}{Phys. Rev. Lett.} \textbf{\bibinfo{volume}{107}},
  \bibinfo{pages}{114302} (\bibinfo{year}{2011}).

\bibitem[{\citenamefont{Kruger et~al.}(2011)\citenamefont{Kruger, Emig, and
  Kardar}}]{Kruger11}
\bibinfo{author}{\bibfnamefont{M.}~\bibnamefont{Kruger}},
  \bibinfo{author}{\bibfnamefont{T.}~\bibnamefont{Emig}}, \bibnamefont{and}
  \bibinfo{author}{\bibfnamefont{M.}~\bibnamefont{Kardar}},
  \bibinfo{journal}{Phys. Rev. Lett.} \textbf{\bibinfo{volume}{106}},
  \bibinfo{pages}{210404} (\bibinfo{year}{2011}).

\bibitem[{\citenamefont{Otey and Fan}(2011)}]{OteyFan11}
\bibinfo{author}{\bibfnamefont{C.}~\bibnamefont{Otey}} \bibnamefont{and}
  \bibinfo{author}{\bibfnamefont{S.}~\bibnamefont{Fan}},
  \bibinfo{journal}{Phys. Rev.~B} \textbf{\bibinfo{volume}{84}}
  (\bibinfo{year}{2011}).

\bibitem[{\citenamefont{Lussange et~al.}(2012)\citenamefont{Lussange,
      Guerout, Rosa, Greffet, Lambrecht, and Reynaud}}]{Lussange12}
  \bibinfo{author}{\bibfnamefont{J.}~\bibnamefont{Lussange}},
  \bibinfo{author}{\bibfnamefont{R.}~\bibnamefont{Guerout}},
  \bibinfo{author}{\bibfnamefont{F.~S.~S.} \bibnamefont{Rosa}},
  \bibinfo{author}{\bibfnamefont{J.~J.} \bibnamefont{Greffet}},
  \bibinfo{author}{\bibfnamefont{A.}~\bibnamefont{Lambrecht}},
  \bibnamefont{and}
  \bibinfo{author}{\bibfnamefont{S.}~\bibnamefont{Reynaud}},
  \bibinfo{journal}{arXiv:1206.0211}  (\bibinfo{year}{2012}).

\bibitem[{\citenamefont{Gu\`{e}rout et~al.}(2012)\citenamefont{Gu\`{e}rout,
  Lussange, Rosa, Hugonin, Dalvit, Greffet, Lambrecht, and
  Reynaud}}]{Guerout12}
\bibinfo{author}{\bibfnamefont{R.}~\bibnamefont{Gu\`{e}rout}},
  \bibinfo{author}{\bibfnamefont{J.}~\bibnamefont{Lussange}},
  \bibinfo{author}{\bibfnamefont{F.~S.~S.} \bibnamefont{Rosa}},
  \bibinfo{author}{\bibfnamefont{J.~P.} \bibnamefont{Hugonin}},
  \bibinfo{author}{\bibfnamefont{D.~A.~R.} \bibnamefont{Dalvit}},
  \bibinfo{author}{\bibfnamefont{J.~J.} \bibnamefont{Greffet}},
  \bibinfo{author}{\bibfnamefont{A.}~\bibnamefont{Lambrecht}},
  \bibnamefont{and} \bibinfo{author}{\bibfnamefont{S.}~\bibnamefont{Reynaud}},
  \bibinfo{journal}{Phys. Rev.~B} \textbf{\bibinfo{volume}{85}}
  \bibinfo{pages}{180301} (\bibinfo{year}{2012}).

\bibitem[{\citenamefont{Rodriguez et~al.}(2012)\citenamefont{Rodriguez, Reid,
  and Johnson}}]{RodriguezReid12}
\bibinfo{author}{\bibfnamefont{A.~W.} \bibnamefont{Rodriguez}},
  \bibinfo{author}{\bibfnamefont{M.~T.~H.} \bibnamefont{Reid}},
  \bibnamefont{and} \bibinfo{author}{\bibfnamefont{S.~G.}
  \bibnamefont{Johnson}}, \bibinfo{journal}{arXiv:1206.1772}
  (\bibinfo{year}{2012}).

\bibitem[{\citenamefont{Levin et~al.}(2010)\citenamefont{Levin, McCauley,
  Rodriguez, Reid, and Johnson}}]{LevinMc10}
\bibinfo{author}{\bibfnamefont{M.}~\bibnamefont{Levin}},
  \bibinfo{author}{\bibfnamefont{A.~P.} \bibnamefont{McCauley}},
  \bibinfo{author}{\bibfnamefont{A.~W.} \bibnamefont{Rodriguez}},
  \bibinfo{author}{\bibfnamefont{M.~T.~H.} \bibnamefont{Reid}},
  \bibnamefont{and} \bibinfo{author}{\bibfnamefont{S.~G.}
  \bibnamefont{Johnson}}, \bibinfo{journal}{Phys. Rev. Lett.}
  \textbf{\bibinfo{volume}{105}}, \bibinfo{pages}{090403}
  (\bibinfo{year}{2010}).

\bibitem[{\citenamefont{McCauley et~al.}(2012)\citenamefont{McCauley, Reid,
  Kruger, and Johnson}}]{McCauleyReid12}
\bibinfo{author}{\bibfnamefont{A.~P.} \bibnamefont{McCauley}},
  \bibinfo{author}{\bibfnamefont{M.~T.~H.} \bibnamefont{Reid}},
  \bibinfo{author}{\bibfnamefont{M.}~\bibnamefont{Kruger}}, \bibnamefont{and}
  \bibinfo{author}{\bibfnamefont{S.~G.} \bibnamefont{Johnson}},
  \bibinfo{journal}{Phys. Rev.~B} \textbf{\bibinfo{volume}{85}},
  \bibinfo{pages}{165104} (\bibinfo{year}{2012}).

\bibitem[{\citenamefont{Bimonte}(2009)}]{bimonte09}
\bibinfo{author}{\bibfnamefont{G.}~\bibnamefont{Bimonte}},
  \bibinfo{journal}{Phys. Rev. A} \textbf{\bibinfo{volume}{80}},
  \bibinfo{pages}{042102} (\bibinfo{year}{2009}).

\bibitem[{\citenamefont{Chapuis et~al.}(2008)\citenamefont{Chapuis, Laroche,
  Volz, and Greffet}}]{Chapuis08prb}
\bibinfo{author}{\bibfnamefont{P.-O.} \bibnamefont{Chapuis}},
  \bibinfo{author}{\bibfnamefont{M.}~\bibnamefont{Laroche}},
  \bibinfo{author}{\bibfnamefont{S.}~\bibnamefont{Volz}}, \bibnamefont{and}
  \bibinfo{author}{\bibfnamefont{J.-J.} \bibnamefont{Greffet}},
  \bibinfo{journal}{Phys. Rev. B} \textbf{\bibinfo{volume}{77}},
  \bibinfo{pages}{125402} (\bibinfo{year}{2008}).

\bibitem[{\citenamefont{Huth et~al.}(2010)\citenamefont{Huth, Ruting, Biehs,
  and Holthaus}}]{Huth10}
\bibinfo{author}{\bibfnamefont{O.}~\bibnamefont{Huth}},
  \bibinfo{author}{\bibfnamefont{F.}~\bibnamefont{Ruting}},
  \bibinfo{author}{\bibfnamefont{S.-A.} \bibnamefont{Biehs}}, \bibnamefont{and}
  \bibinfo{author}{\bibfnamefont{M.}~\bibnamefont{Holthaus}},
  \bibinfo{journal}{Eur. Phys. J. Appl. Phys.} \textbf{\bibinfo{volume}{50}},
  \bibinfo{pages}{10603} (\bibinfo{year}{2010}).

\bibitem[{\citenamefont{Eberlein and Zietal}(2011)}]{Eberlein11}
\bibinfo{author}{\bibfnamefont{C.}~\bibnamefont{Eberlein}} \bibnamefont{and}
  \bibinfo{author}{\bibfnamefont{R.}~\bibnamefont{Zietal}},
  \bibinfo{journal}{Phys. Rev.~A} \textbf{\bibinfo{volume}{83}},
  \bibinfo{pages}{052514} (\bibinfo{year}{2011}).

\bibitem[{\citenamefont{Jackson}(1998)}]{Jackson98}
\bibinfo{author}{\bibfnamefont{J.~D.} \bibnamefont{Jackson}},
  \emph{\bibinfo{title}{Classical Electrodynamics}}
  (\bibinfo{publisher}{Wiley}, \bibinfo{address}{New York},
  \bibinfo{year}{1998}), \bibinfo{edition}{3rd} ed.

\bibitem[{\citenamefont{Duraffourg and Andreucci}(2006)}]{Duraffourg06}
\bibinfo{author}{\bibfnamefont{L.}~\bibnamefont{Duraffourg}} \bibnamefont{and}
  \bibinfo{author}{\bibfnamefont{P.}~\bibnamefont{Andreucci}},
  \bibinfo{journal}{Phys. Lett. A} \textbf{\bibinfo{volume}{359}},
  \bibinfo{pages}{406} (\bibinfo{year}{2006}).

\end{thebibliography}

\end{document}